\def\unredoffs{\voffset=-1.2truecm\hoffset=-0.5truecm}
  \documentstyle[12pt]{article}
  \unredoffs
\catcode`@=11 \@addtoreset{equation}{section} \catcode`@=12

\def\BR{\bf R}
\def\BC{\bf C}
\def\CP{\rm CP}
\begin{document}
\begin{titlepage}
\begin{flushright}
TIFR/TH/95-15\\
{\tt hep-th/9504003}\\
Revised version \\
August 8, 1995
\end{flushright}
\begin{center}
{\large\bf Covariantising the Beltrami equation in W-gravity}\\\bigskip
Suresh Govindarajan\footnote{E-mail:\ suresh@theory.tifr.res.in}\\
{\it Theoretical Physics Group\\
Tata Institute of Fundamental Research\\
Bombay 400 005 INDIA}
\end{center}
\smallskip
\begin{center}
{\it to appear in Letters in Mathematical Physics}
\end{center}
\bigskip
\begin{abstract}
  Recently, certain higher dimensional complex manifolds
  were obtained in \cite{wspace} by associating a higher dimensional
  uniformisation to the generalised Teichm\"uller spaces of Hitchin. The extra
  dimensions are provided by the ``times'' of the generalised KdV
  hierarchy. In this paper, we complete the proof that these manifolds
  provide the analog of superspace for W-gravity and that W-symmetry
  linearises on these spaces. This is done by explicitly constructing
  the relationship between the Beltrami differentials which naturally
  occur in the higher dimensional manifolds and the Beltrami differentials
  which occur in W-gravity. This also resolves an old puzzle regarding the
  relationship between KdV flows and W-diffeomorphisms.
\end{abstract}
\begin{center}
{\it Dedicated to the memory of Claude Itzykson}
\end{center}
\end{titlepage}
\section{Introduction}
W-algebras have provided a unifying ground for diverse topics like
integrable systems, conformal field theory, uniformisation and
2-dimensional gravity. It was originally discovered as a natural
generalisation of the Virasoro algebra by Zamolodchikov\cite{zam} and
implicitly in the work of Drinfeld and Sokolov who obtained
(classical) W-algebras by equipping the ``coefficients'' of first
order matrix differential operators with the second Gelfand-Dikii
Poisson bracket\cite{DS}.  As a result, one can now associate a
W-algebra for every principal embedding of SL$(2)$ in a semi-simple
Lie group.

In the study of two dimensional gravity in the conformal gauge, it is
well known that the generators of the residual reparametrisation
invariance reduce to two copies of the Virasoro algebra. W-gravity can
be defined as a generalisation of reparametrisation invariance such
that in the ``conformal gauge'' one obtains two copies of the
corresponding W-algebra. In W-gravity, the matrix differential
operator is supplemented by another equation which is usually referred
to as the {\it Beltrami equation}. The Beltrami equation is
generically a complicated non-linear equation. These two equations can
be rewritten as flatness conditions of a connection on a semi-simple
group\cite{flat}.

In a different context, Hitchin constructed certain generalised Teichm\"uller
spaces associated to principal embeddings of SL$(2)$ in a semi-simple
Lie group\cite{hitchin}. These Teichm\"uller spaces arose as the
moduli space of solutions of self-duality equations related to certain
stable Higgs bundles. Recently, an explicit relationship has been
obtained between these two constructions\cite{proposal}. This was done
by showing that the equivalence of the Teichm\"uller spaces
constructed by Hitchin to the Teichm\"uller spaces for W-gravity.

Given this rich algebraic structure, it is natural to attempt a
geometric picture for W-algebras. Various aspects of this issue have
been tackled independently by many authors and is universally referred
to as W-geometry\cite{wgeom,flat,proposal}. This geometry is naturally
related to W-gravity. For the case of Virasoro algebra, the geometric
structure has been studied extensively since the times of Poincar\'e and
Klein and is related to the uniformisation of Riemann surfaces.
Recently, a uniformisation in higher dimensions was shown to be
related to the Teichm\"uller spaces constructed by
Hitchin\cite{wspace}.  The extra dimensions are provided by a finite
subset of the ``times'' of the generalised KdV
hierarchy\cite{wspace,gomis}. In order to relate these higher
dimensional manifolds to W-gravity, one has to obtain the relationship
between KdV flows and W-diffeomorphisms. The fact that these two are
not the same was first observed by Di Francesco {\it et al.}\cite{diFr}.

This relationship is addressed in this paper by studying the Beltrami
equation on this higher dimensional setting. The Beltrami
differentials which naturally occur here are related to KdV flows and
not to W-diffeomorphisms.  In this paper, we explicitly construct the
relationship between the two sets of Beltrami differentials. The
construction is based on the fact that the Beltrami differentials of
W-gravity transform as tensors on the Riemann surface, unlike the higher
dimensional Beltrami differentials.

The paper is organised as follows.  In section 2, we discuss the
results of Di Francesco {\it et al.} who covariantised the matrix
differential operator of Drinfeld and Sokolov. In section 3, we
briefly discuss higher dimensional uniformisation and how higher
dimensional generalisations of Riemann surfaces are constructed.  In
section 4, we introduce the Beltrami equation in the higher
dimensional setting and then using techniques described in section 2
to the Beltrami equation, we relate the two sets of Beltrami differentials.

\section{Covariantising the DS equation}

The first order matrix differential equation of Drinfeld and Sokolov
can be converted into an ordinary differential equation of higher
order, or more generally into a system of differential equations (the
DS equation). In this section, we discuss how the DS operators can be
rewritten in a covariant form. This has been done by Di Francesco {\it
et al.} and we will present their results\cite{diFr}. See also related
work\cite{matone}.  For simplicity, we shall concentrate on the case
of $A_{n-1}$ in this paper.  In this case, the DS equation is an
$n$-th order linear differential operator $L$ of the form
\begin{equation}
  L \equiv \partial_z^{n} + u_2(z)\ \partial_z^{n-2} + \cdots
+  u_{n}(z) \quad,
 \label{ediffa} \end{equation}
The differential equation associated with this operator is
\begin{equation}
L\ f =0 \quad.  \label{ediff}
\end{equation}
The vanishing of the coefficient of $\partial_z^{n-1}\ f$ in the above
differential equation implies that the Wronskian of the solutions is a
constant (i.e., independent of the variable $z$). This condition is
consistent across charts on a Riemann surface, provided the Wronskian
is a scalar. This implies that $f$ is a
${{(1-n)}\over2}$-differential. Then the operator $L$ provides a map
from $K^{(1-n)/2}$ to $K^{(1+n)/2}$, where $K$ is the canonical
(holomorphic) line bundle on the Riemann surface.

The form invariance of the operator $L$ under coordinate
transformations enables us to derive the transformation properties of
the the $u_i$. However, they do not transform as nice tensors. Di
Francesco {\it et al.} have worked out an invertible change of
variables $u_i\rightarrow w_i$ such that $w_i$ for $i>2$ transform as
$i$-differentials and $u_2=w_2$ transform like the
Schwarzian\cite{diFr}. This change of variables is non-linear in
$u_2$. We shall summarise their results here. Rather than directly
construct the change of variables it is simpler to rewrite the
differential operator $L$ in terms of covariant operators
$\Delta_k(w_k,u_2)$ for $k=2,\ldots,n$ such that
\begin{equation}
L = \sum_{k=2}^n \Delta_k(w_k,u_2)\quad.
\label{eops}
\end{equation}
Each of these operators $\Delta_k$ provide (covariant) maps from
$K^{(1-n)/2}$ to $K^{(1+n)/2}$. When $u_2=0$, one can write
\begin{equation}
\Delta_k(w_k,u_2=0) = \sum_{i=0}^{n-k}\alpha_{k,i} \
 w_k^{(i)}\  \partial_z^{n-k-i}\quad,
\end{equation}
where $w_k^{(i)}\equiv \partial_z^i w_k$. The covariance of the
operator under M\"obius transformations which preserve the $u_2=0$
condition fixes $\alpha_{k,i}$ to be
\begin{equation}
\alpha_{k,i}=
{{\pmatrix{ k+i-1\cr i}\pmatrix{n-k \cr i}}\over\pmatrix{2k+i-1\cr i}}
\end{equation}

Starting from $u_2=0$, one can make $u_2\neq0$ by means of a
(non-M\"obius) change of coordinate $z\rightarrow z(t)$. Then
$u_2(t)=c_n S[z,t]$ where $c_n=(n^3-n)/12$ and
$S[z,t]={{z'''}\over{z'}} - {3\over2}({{z''}\over{z'}})^2$ (Here $'$
refers to differentiating w.r.t. $t$.) is the
Schwarzian. The $\Delta_k$ transform as
\begin{equation}
\Delta_k(w_k(t),u_2) = J^{(1+n)/2} \sum_{i=0}^{n-k}\alpha_{k,i} \
[(J^{-1} \partial_t)^i J^{-k} w_k] [J^{-1} \partial_t]^{n-k-i}
J^{(n-1)/2} \quad,
  \label{edelk}
\end{equation}
where $J\equiv {{dz}\over{dt}}$. It is a non-trivial fact that the
right hand side of the above expressing depends on $J$ only through
the combination which occurs in the Schwarzian. The proof of this
statement is given in ref. \cite{diFr}. By using the relationship
between the Schwarzian and $u_2$, we obtain the complete $u_2$
dependence of $\Delta_k$.

The $u_i\rightarrow w_i$ transformation can be now explicitly obtained
by comparing the two sides of eqn. (\ref{eops}). One obtains
\begin{equation}
w_k = \sum_{s=0}^{k-2} \beta_{k,s} u_{k-s}^{(s)} + {\rm non-linear\
terms\ involving \ }u_2 \quad,
\end{equation}
where
$$
\beta_{k,s}= (-)^s
{{\pmatrix{ k-1\cr s}\pmatrix{n-k-s \cr s}}\over\pmatrix{2k-2\cr s}}\quad.
$$
We refer the reader to ref. \cite{diFr} for more details.

\section{Higher Dimensional Uniformisation}

In this section, we will briefly describe the construction of higher
dimensional generalisation of Riemann surfaces associated with
PSL$(n,\BR)$.  The reader is referred to \cite{wspace} for more
details.

It is a well known result that Riemann surfaces
of genus $g>1$ can be obtained by the quotient of the upper half plane
(with Poincar\'e metric of constant negative curvature) by a Fuchsian
subgroup of PSL$(2,\BR)$. This is usually referred to as the
Uniformisation Theorem. The space of all Fuchsian groups furnishes the
Teichm\"uller space of Riemann surfaces.  Hitchin generalised these
Teichm\"uller spaces by replacing PSL$(2,\BR)$ with any semi-simple
group\cite{hitchin}. In recent work\cite{wspace}, certain
generalisations of Riemann surfaces were constructed whose
Teichm\"uller spaces are precisely those of Hitchin.

        The method employed in \cite{wspace} is a generalisation of
the differential equation approach to uniformisation as studied by
Poincar\'e and Klein in the 1880's.  Poincar\'e considered the second
order Fuchsian linear differential equation on a Riemann surface
$\Sigma$ of genus $g>1$
\begin{equation}
[\partial_z^2 + u_2(z)]\ f=0\quad,
\end{equation}
where $u_2(z)$ have no singularities\footnote{Regular singular
points correspond to punctures and will not be considered here.}. This
corresponds to $n=2$ in eqn. (\ref{ediff}).

Locally (on a chart), the differential equation has two linearly
independent solutions $f^1$ and $f^2$, which can be considered as
homogeneous coordinates on $\CP^1$.  On analytically continuing the
solutions along any of the cycles of the Riemann surface, the
solutions go into linear combinations of each other. Thus one can
associate a monodromy matrix to each cycle which encodes this
mixing. The analyticity of the solutions implies that this matrix
depends only on the homotopy class of cycle. The set of monodromy
matrices corresponding to each element of $\pi_1(\Sigma)$ form the
monodromy group $\Gamma$ of the differential equation (which is
isomorphic to $\pi_1(\Sigma)$).  Generically, one obtains that the
monodromy group is a subgroup of PSL$(2,\BC)$ (provided we normalise
the basis such that its Wronskian is 1). However, we shall restrict
ourselves to those differential equations which have a monodromy group
$\Gamma$ which is a Fuchsian group and hence a subgroup of
PSL$(2,\BR)$.

The maps from $\Sigma$ to $\CP^1$ are thus not single
valued and hence when one changes charts, there is a PSL$(2,\BR)$
matrix which changes the basis.  Such maps are called {\it
polymorphic} and this multivaluedness encodes the monodromy data of
the differential equation.
Using a standard trick, one can globalise the patch data on the
Riemann surface by lifting the differential equation to the universal
cover of the Riemann surface.  Now the differential equation gives a
map from the universal cover to $\CP^1$. Unlike before, now one
obtains a nice one-to-one map with the monodromy group encoding the
fundamental group of the underlying Riemann surface. When the
monodromy group is Fuchsian, the image of the universal cover in
$\CP^1$ is the upper half plane.  This follows from the fact that under
the action of PSL$(2,\BR)$, $\CP^1$ splits into three parts -- the
upper half plane, the circle and the lower half plane. We can choose
the image to be the upper half plane with no loss of generality. The
monodromy group tesselates the upper half plane with $4g$-gons (whose
sides are geodesics in the Poincar\'e metric). The $4g$-gon represents
a Riemann surface of genus $g$ just as the torus can be represented by
a tesselation of the complex plane by a lattice.  Thus one recovers
the standard case of uniformisation of Riemann surfaces.

We shall now generalise this to the case of $n$-th order Fuchsian
differential equations (again with no regular singular points) on
$\Sigma$ with the monodromy group given by an element of the
Teichm\"uller space of Hitchin corresponding to the group
PSL$(n,\BR)$. Now one gets polymorphic maps from the Riemann surface
to $\CP^{n-1}$. Again, as before, we lift the differential to the
universal cover of $\Sigma$. Let the image of the universal cover be
$\Omega$. $\Omega$ has complex dimension one like the Riemann surface
and the monodromy group tesselates the image as before. However, this
is not sufficient to create the higher dimensional manifold promised
earlier. There exist a set of deformations of the differential
equation which preserve the monodromy group (and hence the point in
the Teichm\"uller space). These are the so called isomonodromic
deformations which are parametrised by the times of the generalised
KdV hierarchy\cite{DS}.
\begin{equation}
{{\partial L}\over{\partial t_i}} = [ (L^{i/n})_+,L]\quad,\label{ekdv}
\end{equation}
where $(L^{i/n})_+$ represents the differential operator part of the
pseudo-differential operator $L^{i/n}$ and $t_1\equiv z$. Further, the
$u_i$ have been extended to be functions of all the times such that
equation (\ref{ekdv}) is satisfied. Since only the first $(n-1)$ times
furnish coordinates in $\CP^{n-1}$, we shall restrict to only those
times.

Now, for each of these times, one obtains the image of the universal
cover in $\CP^{n-1}$, which we shall call $\Omega(\{t_i\})$ for all
$t_i$, $i=2,\ldots,n-1$. Form the union $\tilde\Omega\equiv\cup
\Omega(\{t_i\})$. The monodromy group $\Gamma$ tesselates each of the
time slices and hence all of $\tilde\Omega$. The higher dimensional
manifold, which we shall call the {\it W-manifold} is obtained as
$\tilde\Omega/\Gamma$. By construction, this has complex dimension $(n-1)$.

\section{Covariantising the Beltrami Equation}

In this section, we first introduce the Beltrami equation on the
W-manifold. The Beltrami differentials which occur in
this equation are naturally related to the generalised KdV flows. By
projecting this equation onto the Riemann surface, we show how one can
explicitly construct a relationship to the Beltrami differentials
which occur in W-gravity. This solves the long standing puzzle
regarding the relationship between W-diffeomorphisms and generalised
KdV flows raised by Di Francesco {\it et al.}.

In the previous section, we had restricted ourselves to the holomorphic part
and implicitly assumed that the $f$ were holomorphic. The change of complex
structure on the W-manifold is given by the Beltrami equation\cite{wspace}
\begin{equation}
  [\bar\partial_{\bar z} + \mu_{\bar z}^{\ i} \partial_i -{1\over n}
  (\partial_i\mu_{\bar z}^{\ i}) ]\ f=0\quad,
  \label{ebeltramia}
\end{equation}
where $i=1,\ldots,(n-1)$ and $\mu_i\equiv \mu_{\bar z}^{\ i}$ are the
generalised Beltrami differentials.\footnote{In general, one would
have expected complex deformations with respect to the other complex
coordinates.  In \cite{wspace}, it was however argued that the
W-manifolds are rigid to such complex deformations and hence we will
not consider them.}  On projecting this Beltrami equation to the
Riemann surface, we obtain
\begin{equation}
  [ \bar{\partial}_{\bar z} + \mu_i (L^{i/n})_+ -{1\over n}
  (\partial_i\mu_{i}) ]\ f =0\quad,
  \label{ebeltramib}
\end{equation}
where we have used the relation $\partial_i f = (L^{i/n})_+f $.
However, the $\mu_i$ have complicated transformations under coordinate
transformations on the Riemann surface. We shall derive new Beltrami
differentials $\rho_i$ from the $\mu_i$ such that they transform as
$(-i,1)$ differentials. The $\rho_i$ are precisely the Beltrami differentials
associated with W-gravity.
As we shall explain below, this is the dual
of the $u_i\rightarrow w_i$ transformation. We shall rewrite the
projected Beltrami equation as follows
\begin{equation}
  [\sum_{i=0}^{n-1} {\cal B}_i(\rho_i,w_2) + {\rm terms\ involving\ }
  w_i {\rm \ for\ } i>2 ] \ f =0\quad,
  \label{erewrite}
\end{equation}
where ${\cal B}_0 \equiv\bar{\partial}_{\bar z}$.  ${\cal B}_i$ are
(covariant) differential operators which furnish maps from $K^{(1-n)/2}$
to $K^{(1-n)/2} {\bar K}$ and are constructed from $\rho_i$ and $w_2$.
Equating the coefficients of $\partial_z^i$ for $i=1,\ldots,(n-1)$ in
eqns. (\ref{erewrite}) and ({\ref{ebeltramib}), we obtain the
$\mu_i\rightarrow\rho_i$ transformation. This is an invertible
transformation. Further, we have assumed that this transformation is
independent of $w_i$ for $i>2$. Towards the end of this section we
shall provide evidence that this assumption is valid.

When $w_i=0$, one has $(L^{i/n})_+=\partial_z^i$ and ${\cal B}_i$ can
be written as
\begin{equation}
  {\cal B}_i(\rho_i,w_2=0) \equiv \sum_{s=0}^{i} \gamma_{i,s}
  \rho_i^{(s)} \partial_z^{i-s}\quad,
  \label{ebi}
\end{equation}
with $\gamma_{i,0}=1$ fixing the normalisation of the $\rho_i$.
Comparing equations (\ref{ebeltramib}) and (\ref{ebi}), we obtain
\begin{equation}
\mu_j = \sum_{i\geq j} \gamma_{i,i-j}\rho_i^{(i-j)} + {\rm \ terms
\ involving\ } w_2\quad.
  \label{emurho}
\end{equation}
The unknown coefficients $\gamma_{i,j}$ are determined by requiring
that ${\cal B}_i f$ transforms covariantly under Mobi\"us
transformations which preserve the $w_2=0$ condition.
We obtain
\begin{equation}
\gamma_{i,s} = (-)^s {{\pmatrix{{n+s-i-1}\cr {s}} \pmatrix{{i}\cr
      {s}}}\over   \pmatrix{{2i}\cr {s}}}
  \label{ecoeff}
\end{equation}
Restoring the dependence on $w_2$ is now easy. Under arbitrary
coordinate transformations, the $w_2=0$ condition is not preserved.
Like in the case of $\Delta_k$, all one has to do is to make an
arbitrary coordinate transformation of the ${\cal B}_i(\rho_i, w_2=0)$
and identify the terms corresponding to the Schwarzian of the
transformation with $w_2$. Further, the arguments in ref. \cite{diFr}
showing that $\Delta_k$ depend on $J$ via the combination occuring in
the Schwarzian is also valid for the ${\cal B}_i$. The ${\cal B}_i$
are given by
\begin{equation}
{\cal B}_i(w_k(t),w_2) = J^{(1-n)/2} \sum_{s=0}^{i}\gamma_{i,s} \
[(J^{-1} \partial_t)^s J^{i} \rho_i] [J^{-1} \partial_t]^{i-s}
J^{(n-1)/2} \quad,
  \label{ebib}
\end{equation}

Explicitly, the first few ${\cal B}_i$ (with non-zero $w_2$) are
\begin{eqnarray}
{\cal B}_1 &=& \rho_1 \partial_z -{{n-1}\over2} \rho_1'\quad,\nonumber{}\\
{\cal B}_2 &=& \rho_2 \partial_z^2 - {{n-2}\over2} \rho_2' \partial_z
+ {{(n-1)(n-2)}\over{12}} \rho_2''
+{2\over n}\rho_2 w_2\quad,  \label{enthree}\\
{\cal B}_3 &=&
 \rho_3 \partial_z^3 - {{n-3}\over2} \rho_3' \partial_z^2
+ ({{(n-2)(n-3)}\over{10}} \rho_3''+{{6(3n^2-7)}\over{5(n^3-n)}}
w_2\rho_3) \partial_z\nonumber \\
&-& {{(n-1)(n-2)(n-3)}\over{120}} \rho_3^{(3)}
-{{(4n+7)(n-3)}\over{5n(n+1)}}w_2\rho_3'
 -{{3(n+2)(n-7)}\over{10n(n+1)}}w_2'\rho_3
\nonumber{}
\end{eqnarray}
Finally, by again comparing (\ref{erewrite}) with (\ref{ebeltramib}) with
the $w_2$ also restored in $(L^{i/n})_+$, we obtain the complete
$\mu_i\rightarrow\rho_i$ transformation\footnote{We have used
  $(L^{3/n})_+ = \partial_z^3 + {3\over n}w_2\partial_z+\cdots$ in order to
  obtain the $w_2\rho_3$ term.}.
The first few terms are
\begin{eqnarray}
\mu_1 &=& \rho_1 -  {{n-2}\over2} \rho_2'+({{(n-2)(n-3)}\over{10}}
\rho_3''+{{3(n^2-9)}\over{5(n^3-n)}}w_2\rho_3)+\cdots \nonumber{}\\
\mu_2 &=&\rho_2 - {{n-3}\over2} \rho_3' +\cdots  \label{emurhob}\\
\mu_3&=&\rho_3+\cdots \nonumber{}
\end{eqnarray}
These agree with the known expressions for the change of
variables\cite{wspace,gomis}. Note that in the above expressions, for
a given $n$, only the first $(n-1)$ $\rho_i$ are non-vanishing. Hence
to obtain the required transformation, one has to set the others to zero.

The argument which suggests that only $w_2$ can occur in the
$\mu_i\rightarrow\rho_i$ transformation is as follows. The
$u_i\rightarrow w_i$ transformation is linear for $w_2=0$. The
Beltrami differentials are the conjugate variables to the projective
connections $w_i$ as given by the following natural symplectic
form\footnote{The $(-)^i$ in the definition is essential in order to
agree with the standard definitions of $\rho_i$ and $w_i$. However,
there is a $Z_2$ invariance of the W-algebra given by $w_i\rightarrow
(-)^i w_i$.  So one can absorb the $(-)^i$ into the $w_i$ which would
correspond to a non-standard choice.} on the space
($\rho_i,w_i$)\cite{BFK}
$$
\Omega={1\over{2\pi\hbar}}\sum_{i=2}^{n}\int (-)^i \delta
\rho_{i}\wedge \delta w_{i}\quad.
$$
For $w_2=0$, the change of
variables is nothing but a change of basis.  The
$\mu_i\rightarrow\rho_i$ transformation can then be obtained from the
$u_i\rightarrow w_i$ transformation using the invariance of the
symplectic form $\Omega$. One can check that $\sum(-)^i \mu_{i-1}u_i =
\sum(-)^i \rho_{i-1} w_i$ up to total derivatives using the expressions
derived earlier.  Further, since the non-linear terms in
$u_i\rightarrow w_i$ transformation only involve $w_2$ it seems likely
that this would also be true for the $\mu_i\rightarrow\rho_i$
transformation. Thus, the $w_i$ for $i>2$ cannot occur in the
$\mu_i\rightarrow\rho_i$ transformation.

The relationship between KdV flow and W-diffeormorphisms is provided
in eqn. (\ref{emurhob}) by identifying $\rho_i = {\bar\partial}_{\bar
z}\epsilon^i$ and $\mu_i = {\bar\partial}_{\bar z}\tilde{\epsilon}^i$,
where $\epsilon^i$ parametrise infinitesimal W-diffeomorphisms and
$\tilde{\epsilon}^i$ parametrise infinitesimal KdV flows.  One sees
that $\epsilon^i = \tilde\epsilon^i$ for constant flows provided
$w_2=0$. This resolves the issue raised in \cite{diFr}.  Further, as
conjectured in \cite{wspace} for the case of $W_4$, one can see that
$w_2$ enters the change of variables.

In conclusion, we have presented the relationship between the higher
dimensional manifolds constructed in \cite{wspace} and W-gravity. As a
by product, we have obtained an algorithm to covariantise the Beltrami
equation. Even though we have restricted to the case of $A_n$, the
generalisation to other groups is straightforward.

\noindent{\bf Acknowledgements} This paper is dedicated to the memory
of Claude Itzykson, whose work on W-geometry has been a source of
inspiration to the author.  The author would like to thank M.
Selvadoray for a critical reading of the manuscript.


\end{document}